\documentclass{article}

\usepackage{arxiv}

\usepackage[utf8]{inputenc} % allow utf-8 input
\usepackage[T1]{fontenc}    % use 8-bit T1 fonts
\usepackage{hyperref}       % hyperlinks
\usepackage{url}            % simple URL typesetting
\usepackage{booktabs}       % professional-quality tables
\usepackage{amsfonts}       % blackboard math symbols
\usepackage{nicefrac}       % compact symbols for 1/2, etc.
\usepackage{microtype}      % microtypography
\usepackage{lipsum}
\usepackage{graphicx}
\usepackage{biblatex}
\usepackage{xspace}
\usepackage[most]{tcolorbox}

\usepackage{amsmath}
\usepackage{lipsum}
\usepackage{xcolor, lipsum ,subcaption}
\usepackage{pgfkeys}
\usepackage{textcomp}
\newcommand\blfootnote[1]{%
  \begingroup
  \renewcommand\thefootnote{}\footnote{#1}%
  \addtocounter{footnote}{-1}%
  \endgroup
}

\newcommand{\SODAADV}{\textsc{soda} \textsc{advance}\xspace}
\newcommand{\SODA}{\textsc{soda}\xspace}

\newcommand{\CUPP}{\textsc{cupp}\xspace}
\newcommand{\FORCE}{\textsc{force}\xspace}
\newcommand{\COVERAGE}{\textsc{coverage}\xspace}
\newcommand{\LEET}{\textsc{leet}\xspace}
\newcommand{\CPS}{\textsc{cps}\xspace}

\usepackage{tikz}
\newcommand{\arial}{\small\fontfamily{phv}\selectfont}
\usetikzlibrary{shadows}
\newcommand*\circled[1]{\tikz[baseline=(char.base), every node/.style = {draw, fill=white, drop shadow}]{\node[shape=circle,draw,inner sep=1pt] (char) {\arial #1};}}

\graphicspath{ {./images/} }

\title{Password Strength Analysis Through Social Network Data Exposure: A Combined Approach Relying on Data Reconstruction and Generative Models}

\author{
 Maurizio Atzori \\
 Department of Mathematics and Computer Science \\ University of Cagliari\\ Via Ospedale, 72,
 09124, Cagliari (CA), Italy\\
 \texttt{matzori@unica.it} \\
   \And
 Eleonora Calò\\
  Department of Computer Science\\
  University of Salerno\\ Via Giovanni Paolo II, 132, 84084 Fisciano (SA), Italy\\
  \texttt{ecalo@unisa.it} \\
  \And
 Loredana Caruccio\\
  Department of Computer Science\\
  University of Salerno\\ Via Giovanni Paolo II, 132, 84084 Fisciano (SA), Italy\\
\texttt{lcaruccio@unisa.it}  \\
  \And
  Stefano Cirillo\\
   Department of Computer Science\\
  University of Salerno\\ Via Giovanni Paolo II, 132, 84084 Fisciano (SA), Italy\\
\texttt{scirillo@unisa.it}  \\
\And
Giuseppe Polese\\
Department of Computer Science\\
  University of Salerno\\ Via Giovanni Paolo II, 132, 84084 Fisciano (SA), Italy\\
\texttt{gpolese@unisa.it}
\\
\And
Giandomenico Solimando\\
Department of Computer Science\\
  University of Salerno\\ Via Giovanni Paolo II, 132, 84084 Fisciano (SA), Italy\\
\texttt{gsolimando@unisa.it}
}

\begin{document}
\maketitle
\begin{abstract}
Although passwords remain the primary defense against unauthorized access, users often tend to use passwords that are easy to remember. This behavior significantly increases security risks, also due to the fact that traditional password strength evaluation methods are often inadequate. In this discussion paper, we present \SODAADV, a data reconstruction tool also designed to enhance evaluation processes related to the password strength. In particular, \SODAADV integrates a specialized module aimed at evaluating password strength by leveraging publicly available data from multiple sources, including social media platforms. Moreover, we investigate the capabilities and risks associated with emerging Large Language Models (LLMs) in evaluating and generating passwords, respectively. Experimental assessments conducted with 100 real users demonstrate that LLMs can generate strong and personalized passwords possibly defined according to user profiles. Additionally, LLMs were shown to be effective in evaluating passwords, especially when they can take into account user profile data.\blfootnote{This is a post-peer-review, pre-copyedit version to be published in the Prooceedings of the 33rd Symposium On Advanced Database Systems (SEBD 2025). The final version will be available on \url{CEUR-WS.org}}
\end{abstract}

% keywords can be removed
%\keywords{First keyword \and Second keyword \and More}


\section{Introduction}

Traditional password strength assessments often fall short, as they focus on static syntax rules without considering the semantic context of user choices. Indeed, users generally choose passwords by using keywords easy to remember. However, since much personal information is shared on social networks, attackers can exploit these details to infer user passwords. Thus, through data reconstruction tools, it is possible to reconstruct information semantically related to a context close to users \cite{SODALite}. In this landscape, Large Language Models (LLMs) emerge as both a asset for evaluating password security and a potential threat in generating passwords. This discussion paper examines the privacy risks associated with sharing personal data online and explores the capabilities of LLMs in password evaluation and generation, as proposed in \cite{sodaadv}. The latter presents \SODAADV, an extension of the tool \SODA \cite{SODA}, which includes a new module for evaluating password strength based on information publicly available on social networks. This module exploits some approaches such as \CUPP \cite{cupptool}, \LEET \cite{li2021leet}, \COVERAGE \cite{coverageLi}, and \FORCE \cite{forceEvaluation}, and introduces a new cumulative metric, namely Cumulative Password Strength (\CPS). Furthermore, we present different pipelines, with aim of investigating capabilities and threats associated to the generation and evaluation of passwords by using different LLMs.
The overall evaluation is driven by the following research questions (RQs):
\begin{itemize}
\item[RQ1:] Can we rely on LLMs to suggest complex and easy-to-remember passwords based on publicly available information on social networks?
\item[RQ2:] Can LLMs represent a valid tool to support users in evaluating the strength of passwords based on personal information?
\item[RQ3:] How does the public availability of personal information across multiple social networks impact the capabilities of LLMs to generate and evaluate password strength?
\item[RQ4:] How effective is the prompt-based methodology for password generation and evaluation compared to state-of-the-art models?
\end{itemize} 
\noindent

\section{Combining \SODAADV and LLMs for Evaluating Passwords}\label{sec:sodaAdv_LLMs}

In this section, we describe the \SODAADV tool and the three proposed pipelines that combine the capabilities of LLMs\footnote{\href{www.deepmind.google/technologies/gemini} {www.deepmind.google}, \href{www.chat.openai.com/} {www.chat.openai.com}, \href{www.claude.ai} {www.claude.ai}, \href{www.www.databricks.com} {www.databricks.com}, \href{https://falconllm.tii.ae/falcon.html}{www.falconllm.tii.ae}, and \href{https://llama.meta.com/} {www.llama.meta.com}} (e.g., Google Gemini, ChatGPT, Claude, Dolly, Falcon, and LLaMa) with those of \SODAADV to address password generation and evaluation problems. 

\begin{figure}[t!]
    \centering
    \includegraphics[width=0.7 \textwidth]{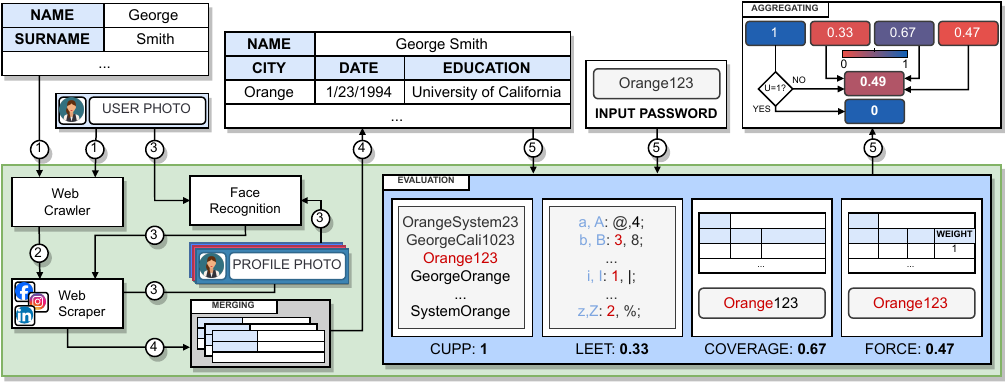}
    \caption{Overview of the modules underlying \SODAADV.}
    \label{fig:Figure1}
\end{figure}

\paragraph{\SODAADV Tool.}
The \SODAADV tool evaluates password strength based on reconstructed personal data from social networks. The \SODAADV pipeline (see Figure \ref{fig:Figure1}) starts with basic user information, i.e., name and photo as input (\circled{1}). It then extracts public data from Facebook, LinkedIn and Instagram using web crawling and scraping techniques (\circled{2}). The tool uses facial recognition to verify the user's identity across all platforms (\circled{3}). Finally, it merges the extracted information (\circled{4}) and evaluates the strength of the provided password based on the reconstructed data (\circled{5}). The evaluation module in \SODAADV uses four methods (i.e., \CUPP, \LEET, \COVERAGE and \FORCE) and a new metric \CPS that combines their results to provide a cumulative value in the range $[0,1]$.

\paragraph{Generation and Evaluation Pipelines.}
The first pipeline is designed to investigate the capabilities of LLMs to generate strong passwords based on specific information provided by users. The process begins with the generation of passwords using LLMs, where each template creates a set of strong but memorable passwords based on user input. The generated passwords are then evaluated with the \SODAADV module, which analyzes their strength. Consequently, each password is labeled as weak or strong according to the strength score.

The second pipeline is designed to investigate the effectiveness of LLMs in assessing the strength of passwords by also considering their semantics in relation to user data. The process begins by generating strong passwords using the best LLM of the Generation Pipeline. Simultaneously, weak passwords are generated by using \CUPP. Once passwords are created, a new prompt is generated to evaluate their strength. The evaluation involves submitting the user data along with the generated passwords to an LLM, which then assigns each password a numerical strength score. Finally, passwords are categorized as weak or strong according to obtained score. Details concerning the above-described pipelines can be found in \cite{sodaadv}.

\paragraph{Data Reconstruction and Password Evaluation Pipeline.}\label{sec:ThirdPipe}

The third pipeline combines the password strength evaluation of \SODAADV with that of the LLMs, using new automated prompting functions for evaluating passwords. They directly consider within a prompt both data reconstructed from the social networks and the results achieved by \SODAADV.
\begin{figure}[t!]
    \centering
    \includegraphics[width=0.7 \textwidth]{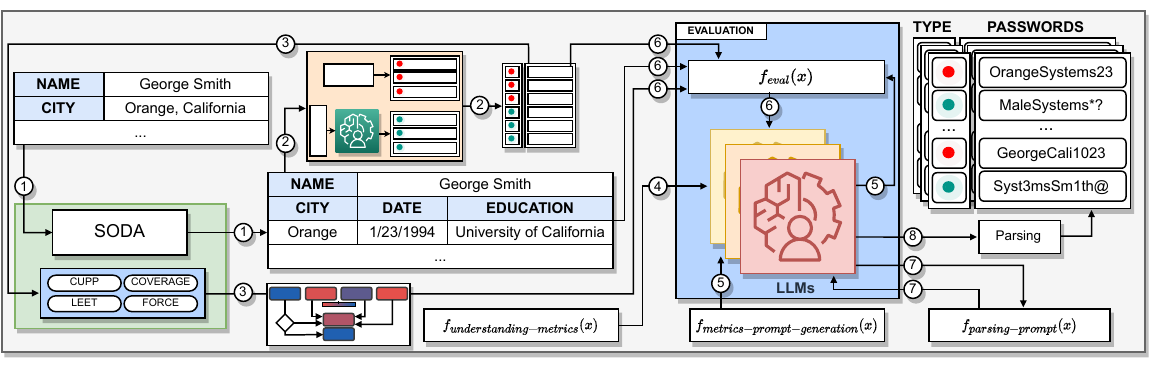}
    \caption{Overview of the data reconstruction and password evaluation pipeline.}
    \label{fig:Figure2}
\end{figure}
As shown in Figure \ref{fig:Figure2}, starting from a small set of user information, we used \SODAADV to reconstruct it using the publicly available information shared on social networks \circled{1}. Then, in \circled{2}, the reconstructed information is used to create a dataset containing both strong and weak passwords associated with the user. In \circled{3}, the set of user passwords is provided to \SODAADV that is responsible for their first evaluation.
Before proceeding with the evaluation step, in \circled{4}, new prompt containing the explanation of each metrics adopted by \SODAADV is provided to LLMs. Moreover, for each of them, in \circled{5}, a new prompt  considered both values resulting from \SODAADV and the data reconstructed from the social networks, is automatically generated, which is then submitted to LM together with the passwords to be evaluated in \circled{6}, each prompt is filled with the user reconstructed data and the evaluation results from \SODAADV, and it is submitted to an LLM together with the passwords to be evaluated. Finally, in \circled{7} and \circled{8}, each password is associated a strength score to identify its category: strong or weak.

\paragraph{Prompt Engineering Approach for Password Strength problem.}

The process of generating passwords required the definition of an ad-hoc prompting function, namely $f_{\text{password-generation}}$, as shown in the following.

\begin{tcolorbox}[enhanced, width=.99\linewidth, center upper, halign=left, left=1mm, right=1mm, floatplacement=!t, fontupper=\small, drop shadow southwest, sharp corners, segmentation style={solid, darkgray!60}]
\begin{center}
\scriptsize
\textbf{On the basis of the following personal information: [Name: George], [Surname: Smith], [City: Orange, California], [Date: 10/23/1994].
Could you generate a set of passwords that do not have to directly contain personal data, but must be easy for the user to memorize?}
\end{center}
\end{tcolorbox}

Instead, the process behind the password evaluation pipeline requires interacting with LLMs at several steps. Among these, we defined a new function $f_{\text{prompt-generation}}$ to ask each LLM to automatically generate prompts for password evaluation and a new function $f_{\text{parsing-prompt}}$ to ask each LLM to provide a strength score for each textual description. These prompts enabled us to automatically create a new prompting function, namely $f_{\text{evaluate-password}}$ for each LLM involved in our study. An example of the prompt automatically generated for ChatGPT follows: 
\begin{tcolorbox}
[enhanced, width=.99\linewidth, center upper, halign=left, left=1mm, right=1mm, floatplacement=!t, fontupper=\small, drop shadow southwest, sharp corners, segmentation style={solid, darkgray!60}]

\centering
\scriptsize
        \textbf{User information: [Name: George], [Surname: Smith], [City: Orange, California], [Date: 10/23/1994], [Education: University of California].
            For each line containing a password that I could use for a social network account, give me an answer for each of them and write whether the password can be considered secure or not, giving secure or not secure. Assess the password's strength using the information supplied by the user, considering factors like its length and ability to resist guessing techniques. 
            Passwords: [OrangeSystems23], [MaleSystems*?], [GeorgeCali1023], [C@liforn1Sm1th49], [Syst3msSm1th@], [0r@nge@n3@]}
\noindent
\end{tcolorbox}

In the third pipeline the interaction with LLMs to evaluate the password strength has required the use of some of the previous prompting functions, and the definition of new ones to explain the metrics (i.e., $f_{\text{understanding-metrics}}$) to LLMs and evaluate the password, by also considering the results of \SODAADV. We manually defined two new prompting functions following the \textit{Manual Template Engineering} strategy \cite{liu2023pre}, and we automatically generated those to evaluate passwords for each LLMs, by means of $f_{\text{metrics-prompt-generation}}$ function. Starting from the generated function $f_{\text{eval}}$, we automatically generate a new specific prompt is provided to each LLM. The prompt generated by ChatGPT is shown below:
\begin{tcolorbox}
[enhanced, width=.99\linewidth, center upper, halign=left, left=2mm, right=2mm, floatplacement=!t, fontupper=\small, drop shadow southwest, sharp corners, segmentation style={solid, darkgray!60}]
\centering
\scriptsize
            \textbf{User information: [Name: George, Surname: Smith, City: Orange, California, Date: 10/23/1994]\\}   
            \textbf{Passwords Evaluation Results:\\}
\textbf{            
               Password; Force; Leet; Coverage; CUPP; CPS \\
                    OrangeSystems; 23; 57; 57; 0; 0.45\\
                    MaleSystems*?; 27; 2; 71; 1; 1\\
                    GeorgeCali1023; 63; 12; 76; 0; 0.50\\
                    C@liforn1Sm1th49; 65; 0; 83; 0; 0.49\\
                   }
            \textbf{Please assess the security of each password listed. Using the user information provided, analyze the password strength based on the following methods: Leet Coverage, Force, CUPP, and Cumulative Password Strength. Upon evaluation, please provide a response of} Strong \textbf{if the password is deemed sufficiently strong and effectively safeguards the user's information based on the provided data, or} Weak \textbf{if the password could potentially be compromised or guessed based on the available details.}   
\noindent
\end{tcolorbox}

The prompts generated by LLMs for evaluating password strength have showed similarities in their structures but have demonstrated differences in formatting and language style. In the following sections, we will show a case study involving real users that allows us to investigate the capabilities of \SODAADV and LLMs to evaluate password strength.

\section{Experimental Evaluation}\label{sec:experimentalevaluation}
The experiments in this study aim to evaluate how password strength can be affected by the information publicly available on social network platforms from both syntactical and semantic perspectives. To this end, we investigate the behavior of \SODAADV and generative LLMs following the three different pipelines discussed in the previous section. We involved $100$ users, each of whom filled out an information survey and an authorization form for profiling their social network using \SODAADV. Among the questions submitted to users, we required their name, surname, and a photo. The collected data is used as starting points of the evaluation. Notice that, we obtained the explicit consent by users, in compliance with GDPR \cite{GDPR}.

\paragraph{Technical Settings.}\SODAADV was implemented using Python version 3.10.2 on the server side and using web programming frameworks for graphical interfaces. Concerning LLMs, we adopt ChatGPT 3.5.5, Claude 2.1, LLaMa 2024.2.19.1, Falcon in its version at 40B, Google Gemini 1.0, and Dolly-v2-12b. Moreover, for the analysis of the characteristics of the generated passwords we used two different tools (i.e., \textit{Passat} and \textit{Node-password-analyzer})\footnote{\href{www.github.com/HynekPetrak/passat}{www.github.com/HynekPetrak/passat}, \href{www.github.com/T-PWK/node-password-analyzer}{www.github.com/T-PWK/node-password-analyzer}}. Furthermore, to make a comparative evaluation with \SODAADV, we use the \textit{Zxcvbn library} \cite{zxcvbn} in its version $4.4.2$, the \textit{CKL\_PSM library} \cite{ckl}, and the \textit{Semantic PCFG} \cite{SemanticPCFG} tool. The latter tool was trained on plain text passwords extracted from the \textit{Evite}\footnote{\href{https://haveibeenpwned.com}{www.haveibeenpwned.com}} dataset. Finally, for generative password comparison, we use the \textit{PassBERT model} \cite{passbert}.

\paragraph{RQ1.}\label{sec:RQ1}
The characteristics of the generated passwords revealed that each LLM exhibits distinct patterns in the generation of strong passwords, with variations in syntactical complexity and the combination of letters/characters. Thus, to evaluate the strength of passwords we used the new metric \CPS of \SODAADV. In average, we obtained that Claude, Google Gemini, and ChatGPT outperform the other LLMs achieving a score of $0.82$, $0.75$, and $0.74$, respectively. On the other hand, Dolly, LLaMa, and Falcon have generated more weak passwords, achieving a score of $0.65$, $0.66$, and $0.66$, respectively. This is probably due to their tendency to generate repetitive or predictable passwords, using recurring and easily guessable patterns. 

\paragraph{RQ2.}\label{sec:RQ2} 
Starting from the values provide by \CPS, we consider a password as \textit{strong} when its strength score is greater than or equal to $0.55$, \textit{weak} otherwise. Those, we are able to get a binary evaluation of passwords and compare the results achieved by LLMs with those achieved by methods proposed in the state-of-the-art. By considering the average value achieved by each LLM, Claude obtained the highest values for accuracy, precision, recall, and F1-score, i.e., $0.75$, $0.76$, $0.75$, and $0.75$, respectively. The high precision score indicates that it has a low rate of False Positive, meaning that it correctly identifies strong passwords with a high degree of confidence. To further investigate if the ensemble of different LLMs improves the values of metrics, we considered two different ensembles: $i$) including all the LLMs and $ii$) including the three LLMs with the highest scores; but both performed lower than Claude.

\paragraph{RQ3.} By combining social media data with the semantic capabilities of LLMs, password strength evaluations significantly improved with respect to scenario in which a few user data is provided to LLMs. Compared to the latter scenario, the inclusion of broader personal information led to better performance across most models. For instance, Falcon improved its precision from $0.48$ to $0.77$ and ChatGPT reached high scores in accuracy, precision, recall, and F1-score. Instead, Claude showed the best overall performance (i.e., accuracy equal to $0.77$ and precision equal to $0.89$). Ensemble models also benefited, likely due to the enhanced performance of individual LLMs. These improvements suggest that public social media data provides valuable context, allowing LLMs to make more accurate assessments. However, this also raises privacy concerns: as more personal data becomes accessible, users face increased risks. LLMs could be exploited by attackers to guess passwords based on publicly shared information. This highlights the importance of strong privacy settings, secure password practices, and the need for clear ethical and legal guidelines regarding the use of LLMs.

\paragraph{RQ4.}\label{sec:RQ4}
To evaluate the capabilities of LLMs in both password generation and evaluation tasks, as well as the effectiveness of \SODAADV in assessing password strength, we analyzed the medium-security passwords and compared the results with state-of-the-art tools.

\textit{\underline{Medium Password Strength evaluation.}} Starting from the initial dataset provided by 100 users, we generated a set of $30$ passwords for each user using the prompt $f_{\text{password-generation}}$. The values of \CPS obtained for medium-strength passwords generated by LLMs and evaluated through \SODAADV range between $0.36$ and $0.60$. In particular, Claude, Google Gemini, and ChatGPT outperform all other LLMs achieving the highest number of medium passwords. Then, to assess the evaluation capabilities of LLMs and \SODAADV, we asked each model to evaluate each password. By using the evaluation pipeline, the classification task involving multiple labels (i.e., \textit{weak}, \textit{medium}, \textit{strong}), significantly reduced the performance of all LLMs with respect to the binary classification task (i.e., \textit{weak} and \textit{strong}). In particular, we have noticed that most of the passwords correctly evaluated were weak passwords, containing recurrent patterns and combinations of user data. Instead, LLMs were not able to discriminate passwords between strong and medium levels. Conversely, with the data reconstruction and password evaluation pipeline, the overall performance was higher, demonstrating that Claude outperformed all other LLMs. Our analysis shows that the initial evaluation provided by \SODAADV effectively supported LLMs in distinguishing between weak, medium, and strong passwords. 

\begin{figure}[t!]
    \centering
    \includegraphics[width=0.30\textwidth]{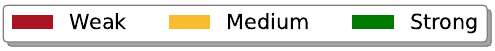}  \\
    \includegraphics[width=0.8\textwidth]{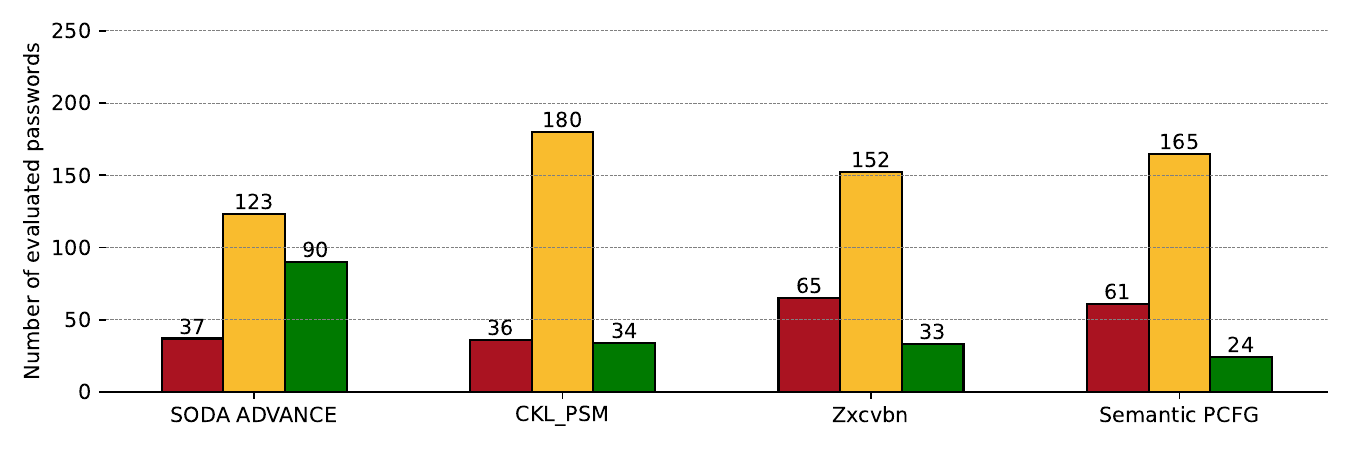}
    \caption{Result obtained from \SODAADV, CKL\_PSM, Zxcvbn, and Semantic PCFG on \textit{Evite} dataset.}
    \label{fig:Figure3}
\end{figure}

\textit{\underline{Comparative evaluation with state-of-the-art tools.}} We performed a comparison with \SODAADV and some of the most recent tools for password evaluation available in the state-of-the-art, i.e., Zxcvbn, CKL\_PSM, and Semantic PCFG. In order to be able to compare the values obtained from the library and tools with those of the evaluation module of \SODAADV, we uniform the ranges to fit the strength of the passwords in three categories, \textit{weak}, \textit{medium}, and \textit{strong}. For the purposes of our evaluation, we extracted a random sample of $250$ passwords, ranging in length from $8$ to $25$ characters. Figure \ref{fig:Figure3} shows the results of  \SODAADV, CKL\_PSM, Zxcvbn, and Semantic PCFG on the considered set of passwords. As we can see, most of the passwords have been classified as medium by all tools, and only a few of them as strong. \SODAADV has demonstrated good capabilities of evaluation for the passwords containing these types of information, classifying them as weak. Moreover, \SODAADV classified as medium some passwords consisting of simple dictionary words not semantically linked to users. These types of passwords have been considered strong by the methods that evaluate these attempts, i.e., CKL\_PSM, Zxcvbn, and Semantic PCFG, since they have a medium-complex syntax that requires a large number of attempts to crack. This is probably due to the metrics for the analysis of syntax included in the \CPS.

By summarizing, we have noticed that no model excels at evaluating password strength. As we expected, \SODAADV demonstrated good evaluation capabilities for passwords that contain some user information but overestimates the complexity of passwords when they contain words not semantically linked to the user. On the other hand, tools that evaluate passwords based on crack attempts often underestimate the strength of passwords with complex syntax if they contain information related to the user. However, as also demonstrated for LLMs, considering the problem of evaluating password strength based on semantics with three levels of strength is extremely more difficult and the evaluations are less accurate. 

\textit{\underline{Evaluating passwords with a state-of-the-art model.}} To further investigate the password-generation capabilities of LLMs, we evaluated the strength of the passwords with PassBERT \cite{passbert}, which is one of the most recent models in the literature for making focused attacks on passwords. PassBERT uses the fine-tuning paradigm for password-guessing attacks, with a pre-trained password model and different fine-tuning approaches. Among them, we considered Targeted Password Guessing (TPG) which aims to estimate the number of guesses of cracking the input password given a set of leaked passwords. For the purposes of our evaluation, we considered $100$ users and their $250$ strong passwords generated by LLMs. Moreover, we considered the weak passwords inferred by \CUPP as leaked passwords. For each strong password, we evaluated its strength with the PassBERT model and the TPG approach. By considering $250$ passwords for each user, we collected a total of $25,000$ strong passwords. The results showed that among the strong passwords, only the passwords of a small set of users were inferred by PassBERT. Specifically, PassBERT was able to identify only $22$ passwords out of the $25,000$ evaluated, probably due to the complexity of the syntax of these passwords. In fact, although the passwords generated by LLMs are based on personal information about the user and therefore easy to remember, they are also syntactically complex and difficult to crack for models such as TPG. These results, together with those achieved from the previous evaluation, underscore the robustness of using LLMs for generating secure passwords semantically related to the information of the users and highlight the limited effectiveness of an advanced targeted guessing model, i.e., PassBERT.

\section{Conclusion and Future Directions}\label{sec:conclusion}
We have investigated the threats related to the definition of password when users publicly share their data on social network platforms. To this end, we have first proposed a new data reconstruction tool, namely \SODAADV, capable of reconstructing public user data and evaluating a password according to them. Moreover, we have designed three different pipelines aiming to evaluate the performance of emerging LLMs, in the generation of strong passwords and the evaluation of their strength by a new ad-hoc prompting functions based on automatic and manual prompt engineering approaches. The experimental evaluations with real users have shown that Claude revealed good capabilities in generating strong passwords and evaluating password strength based on user data. Moreover, the combination of LLMs with the \SODAADV tool has led to significant improvements in the password evaluation process with LLMs. To further investigate the effectiveness of LLMs and \SODAADV in password generation and evaluation, we compared it with state-of-the-art approaches. The results highlight that LLMs do not perform well in the generation of medium-level passwords. Instead, the evaluation methods included in \SODAADV performed better in this task. Finally, it has been shown that a very small percentage of strong passwords generated by LLMs succeed in being leaked by PassBERT's TPG model.

The methodologies and results obtained in this study open the research in several new directions. Future research could investigate in-depth the understanding and mitigation of threats, including exploring alternative approaches to password management and authentication in the context of widespread public data availability. In addition, further investigation could focus on enhancing the capabilities of the data reconstruction tool to extract a large set of public information from other Web platforms. Moreover, password strength assessment can be further explored using LLM by investigating the effectiveness of models trained specifically for this problem. Finally, emerging trends related to LLMs require further investigation for a better understanding of how these models treat personal information and whether they comply with European and global regulations.

\section*{Acknowledgments}
{\small This work was partially supported by project SERICS (PE00000014) under the NRRP MUR program funded by the EU - NGEU.}

\printbibliography
%\bibliography{references}  %%% Remove comment to use the external .bib file (using bibtex).
%%% and comment out the ``thebibliography'' section.

\end{document}